\documentclass[aps,showkeys,showpacs,superscriptaddress,a4paper,twocolumn,10pt]{revtex4-2}

\usepackage{amssymb}
\usepackage{amsmath}
\usepackage[dvips]{graphicx}
\usepackage[T2A]{fontenc}
\usepackage{graphicx}
\usepackage[dvips]{color}
\usepackage[breaklinks=true,colorlinks=true,linkcolor=blue,urlcolor=blue,citecolor=blue]{hyperref}
\usepackage{subcaption}
\let\doi\relax
\usepackage{doi}

\begin{document}

\title[Order-parameter-dependent mobility in the convective-viscous Cahn-Hilliard equation]{Order-parameter-dependent mobility \\ in the convective-viscous Cahn-Hilliard equation}

\author{P.~O.~Mchedlov-Petrosyan}
\author{L.~N.~Davydov}
\email[Corresponding author: ]{ldavydov@kipt.kharkov.ua}

\affiliation{National Science Center Kharkiv Institute of Physics and Technology, \\
1, Akademichna St., Kharkiv, 61108, Ukraine}

\author{O.~A.~Osmayev}
\affiliation{National Science Center Kharkiv Institute of Physics and Technology, \\
1, Akademichna St., Kharkiv, 61108, Ukraine}
\affiliation{Ukrainian State University of Railway Transport, 7, Feuerbach Sq., Kharkiv, 61050, Ukraine}

\begin{abstract}
Generally the mobility in the Cahn-Hilliard equation may depend on the order parameter; this was evident already from the original derivation. However, explicit introduction of such dependence results in tremendous calculation difficulties, so the constant mobility approximation was commonly used. On the other hand, in several cases this dependence appeared to be crucial, so more realistic expressions were applied, usually positive powers or polynomials; still, for such dependencies only approximate and/or numerical solution exist. We consider 'reciprocal' linear and quadratic dependencies of mobility on the order parameter; for these dependencies exact traveling wave solutions are obtained. Even more, in a limited interval of the parameters reciprocal quadratic mobilities are rather good approximations to polynomial mobilities.
\end{abstract}


\maketitle

\renewcommand{\theequation}{\arabic{section}.\arabic{equation}}
\section{Introduction}\label{s1}

The present work is devoted to the study of the modified nonlinear convective-viscous Cahn-Hilliard equation with non-constant mobility:
\begin{eqnarray} \label{1.1} & & {\frac{\partial w}{\partial \bar{t}}-2\bar{\alpha }w\frac{\partial w}{\partial \bar{x}} =\frac{\partial }{\partial \bar{x}} \left\{\bar{M}\left(w\right)\right. } \\ & &   {\times\left.\frac{\partial }{\partial \bar{x}} \left[\rho \left(w\!-\!\bar{a}_{1} \right)\left(w\!-\!\bar{a}_{2} \right)\left(w\!-\!\bar{a}_{3} \right)\!-\!\varepsilon ^{2} \frac{\partial ^{2} w}{\partial \bar{x}^{2} }\! +\!\bar{\eta }\frac{\partial w}{\partial \bar{t}} \right]\right\}} . \nonumber \end{eqnarray}
Here $w$ is the order parameter, $\bar{a}_{1} <\bar{a}_{2} <\bar{a}_{3} $ are values of the order parameter, corresponding to two minima and intermediate maximum of the fourth-power polynomial in the thermodynamic potential; $\varepsilon $ is usually presumed to be proportional to the capillarity length, and $\bar{\eta }$ is the viscosity. We first consider three possible expressions for the mobility $\bar{M}\left(w\right)$,
\begin{equation} \label{1.2} \bar{M}_{i} \left(w\right)=\frac{m}{\theta \left(w-\bar{a}_{i} \right)^{2} +\xi \bar{a}_{3}^{2} } ,\, \, \, \, i=1,3 ,  \end{equation}
\begin{equation} \label{1.3} \bar{M}_{s} \left(w\right)=\frac{m}{\theta \left(w-\bar{a}_{1} \right)\left(w-\bar{a}_{3} \right)+\xi \bar{a}_{3}^{2} }  .  \end{equation}
Here $m$ is a dimensional parameter; $\theta >0$ and $\xi >0$ are nondimensional parameters.  When the order parameter $w$ increases in the interval $\left(\bar{a}_{1} ,\bar{a}_{3} \right)$ the mobility $M_{1} $ monotonically decreases, and the mobility $M_{3} $ monotonically increases, see \eqref{1.2}. On the other hand, the mobility $\bar{M}_{s} $ has equal minimal values $\frac{m}{\xi \bar{a}_{3}^{2} } $ at the stationary states $\bar{a}_{1} ,\bar{a}_{3} $ and the maximum value,
\begin{equation} \label{1.4)} \max \left[\bar{M}_{s} \right]=\frac{4m}{4\xi \bar{a}_{3}^{2} -\theta \left(\bar{a}_{3} -\bar{a}_{1} \right)^{2} } , \end{equation}
at the intermediate state $w_{m} =\frac{1}{2} \left(\bar{a}_{1} +\bar{a}_{3} \right)$.

Naturally, to avoid the singularity and negativity of the mobility parameter $\xi $ must be greater than $\xi_{s} $, where
\begin{equation} \label{1.5)} \xi _{s} =\theta \frac{1}{4\bar{a}_{3}^{2} } \left(\bar{a}_{3} -\bar{a}_{1} \right)^{2} . \end{equation}
Eq. \eqref{1.1} with the mobility given by Eqs. \eqref{1.2}-\eqref{1.3} is the modification of the convective-viscous Cahn-Hilliard equation \cite{1,2}. The overview of the history and existing modifications of the Cahn-Hilliard equation (herein we will use the abbreviation CH equation) was given in \cite{2} and will not be repeated here. We also refer to the pioneering papers \cite{3,4} and excellent reviews \cite{5,6}. The CH equation is now a well-established model in the theory of phase transitions as well as in several other fields. The basic underlying idea of this model is that for inhomogeneous system, e.g., system undergoing a phase transition, the thermodynamic potential (e.g. free energy) should depend not only on the order parameter $w$, but on its gradient as well. For the inhomogeneous system the local chemical potential $\bar{\mu }$ is defined as variational derivative of the thermodynamic potential functional. If the thermodynamic potential is the simplest symmetric -- quadratic -- function of gradient this leads to the local chemical potential $\bar{\mu }$ which depends on Laplacian, or for the one-dimensional case -- on the second order derivative of the order parameter $w$. The diffusion flux $J$ is proportional to the gradient of chemical potential $\nabla \bar{\mu }$; the proportionality coefficient is called mobility. With such expression for the flux the diffusion equation, instead of usual second order equation, becomes a forth-order PDE for the order parameter $w$.

The classic CH equation was introduced as early as 1958 \cite{3,4}; the stationary solutions were considered, the linearized version was treated and the instability of homogeneous state identified. However, intensive study of the fully nonlinear form of this equation was started by the seminal paper of Novick-Cohen and Segel \cite{7}. In this paper several aspects of the nonlinear CH equation were discussed, with special attention to the non-constant mobility. This work was continued in \cite{8,9}. Now an impressive amount of work is done on nonlinear CH equation\textbf{\textit{,}} as well as on its numerous modifications, see \cite{5,6}. An important modification was done by Novick-Cohen \cite{10}. Taking into account the dissipation effects which are neglected in the derivation of the classic CH equation, she introduced the \textbf{\textit{viscous}} Cahn-Hilliard equation,
\begin{equation} \label{1.6} \frac{\partial w}{\partial \bar{t}} =\nabla \left[\bar{M}\nabla \left(\bar{\mu }+\bar{\eta }\frac{\partial w}{\partial \bar{t}} \right)\right] ,  \end{equation}
where the coefficient $\bar{\eta }$ is called viscosity; also see \cite{11}. Later several authors considered the nonlinear \textbf{\textit{convective}} CH equation in one space dimension \cite{12,13,14}. In \cite{12,13,14,15,16} several approximate solutions and exact static kink and anti-kink solutions were obtained. The `coarsening' of domains separated by kinks and anti-kinks was also discussed.

To study the joint effects of nonlinear convection and viscosity, Witelski \cite{1} introduced the convective-viscous-Cahn--Hilliard equation with a general symmetric double-well potential $\Phi \left(w\right)$:
\begin{equation} \label{1.7} \frac{\partial w}{\partial \bar{t}} -\bar{\alpha }w\frac{\partial w}{\partial \bar{x}} =\frac{\partial }{\partial \bar{x}} \left[\bar{M}\frac{\partial }{\partial \bar{x}} \left(\bar{\mu }+\bar{\eta }\frac{\partial w}{\partial \bar{t}} \right)\right] ,  \end{equation}
\begin{equation} \label{1.8)} \bar{\mu }=-\varepsilon ^{2} \frac{\partial ^{2} w}{\partial \bar{x}^{2} } +\frac{d\Phi \left(w\right)}{dw} . \end{equation}

With a constraint imposed on nonlinearity and viscosity, the approximate traveling wave solutions were obtained. In \cite{2} for Eq.\eqref{1.7} with polynomial potential and the balance between the applied field and viscosity several exact single- and two-wave solutions were obtained.

As a rule, the mobility $\bar{M}$ may depend on the order parameter. This was evident already from the original derivation \cite{4} and was also discussed in \cite{17}. However, to the best of our knowledge, the systematic mathematical consideration of the order-parameter-dependent mobility in the CH equation was pioneered by Novick-Cohen \cite{9,10}. In this work the stability of the spatially homogeneous state was explored using the energy method. The CH equation with non-constant mobility was discussed in \cite{18,19} studying the morphological changes due to surface diffusion. Later the concentration dependence of mobility was found to be crucial in modeling the motion of interfaces, separating different phases of an alloy (`geometric motion') \cite{20}. In \cite{21,22} concentration-dependent mobility was considered with different functional forms of the thermodynamic potential.

In the study of the later stage of phase separation (coarsening) \cite{23,24} it was found that dependence, or independence of the mobility on the concentration, manifests itself in different forms of asymptotic scaling. The viscous Cahn-Hilliard equation with non-constant mobility was considered in \cite{25}. Furthermore, in this study, the coefficient at the $\left(\nabla w\right)^{2} $ term in the free energy expression was order-parameter-dependent.

Another line of modifications of the Cahn-Hilliard equation with non-constant mobility is the introduction of additional terms, either into the equation, or in the expression for the chemical potential. Schimperna \cite{26} considered both the CH and viscous CH equation with the non-constant mobility and a source term in the chemical potential. Usually, in the above mentioned papers the mobility was a positive power or polynomial of the order parameter. On the other hand, Fujita noticed in \cite{27,28} that for some systems the proper dependence of the diffusion coefficient is a negative power, or even the negative power of a quadratic polynomial. Such dependence appeared also in the Mullins' theory of thermal grooving \cite{29}, see \cite{30,31}, and in other solid-state contexts \cite{32}. The $w$-dependence of the form \eqref{1.2} attracted much mathematical interest when it was discovered that the diffusion equation with this nonlinearity is the single exactly linearisable nonlinear diffusion equation \cite{33,34}.

Our model \eqref{1.1}, \eqref{1.2} combines several features appeared in the above mentioned modifications of the CH equation: convective terms, dissipation and non-constant mobility.
In the absence of the convective term, i.e., for the viscous CH equation we consider additionally two different types of mobility $\bar{M}\left(w\right)=\bar{K}_{i} \left(w\right)$,
\begin{equation} \label{1.9} \bar{K}_{1} \left(w\right)=\frac{m}{\left(w-\bar{a}_{1} \right)}  ,  \end{equation}
\begin{equation} \label{1.10} \bar{K}_{3} \left(w\right)=\frac{m}{\left(\bar{a}_{3} -w\right)} .  \end{equation}

This dependence of the diffusion coefficient was first considered by Fujita \cite{27}. Such type of concentration dependence was observed in several systems, for instance in the ion exchange kinetics \cite{35}, in thin liquid film dynamics \cite{36}, in Boltzman kinetic model \cite{37}, for the fluctuations dynamic in the self-organized systems \cite{38,39}. The diffusion equation with such dependence possesses some special features, which attracted much mathematical interest \cite{40,41,42,43,44,45}.

We first obtain exact traveling wave solutions; then using this example, we examine the effect of the non-constancy of mobility on the solution's parameters, i.e., on the velocity, amplitude and steepness of the front.

\setcounter{equation}{0}
\section{Traveling wave solution}\label{s2}

Introducing the non-dimensional order parameter $u=\frac{w}{\bar{a}_{3} } ,$, non-dimensional coordinate $x=\frac{\bar{x}}{X} ,\, \, X=\frac{\varepsilon }{\bar{a}_{3} \sqrt{\rho } } $, non-dimensional time $t=\frac{\bar{t}}{T} $, $T=\frac{\varepsilon ^{2} }{m\rho ^{2} \bar{a}_{3}^{2} } $, and non-dimensional mobility $M_{i,s} =\frac{\bar{M}_{i,s} \bar{a}_{3}^{2} }{m} $ we rewrite Eqs. \eqref{1.1}-\eqref{1.3} in the non-dimensional form
\begin{eqnarray} \label{2.1} & & \frac{\partial u}{\partial t} -2\alpha u\frac{\partial u}{\partial x} =\frac{\partial }{\partial x} \left\{M_{i,s} \right. \\ & &\times \left.\frac{\partial }{\partial x} \left[\left(u-a_{1} \right)\left(u-a_{2} \right)\left(u-a_{3} \right)-\frac{\partial ^{2} u}{\partial x^{2} } +\eta \frac{\partial u}{\partial t} \right]\right\}, \nonumber \end{eqnarray}
\begin{equation} \label{2.2} M_{i} =\frac{1}{\theta \left(u-a_{i} \right)^{2} +\xi } ,\, \, \, i=1,3 , \end{equation}
\begin{equation} \label{2.3} M_{s} =\frac{1}{\theta \left(u-a_{1} \right)\left(u-a_{3} \right)+\xi } . \end{equation}
Here $a_{3} =1$, however we keep it for the unification of expressions; also we introduced the following notations,
\begin{eqnarray} \label{2.4)} & & \alpha =\frac{T\bar{\alpha }\bar{a}_{3} }{X} =\bar{\alpha }\frac{\varepsilon }{m} \rho ^{-\frac{3}{2} } ;\, \, a_{1} =\frac{\bar{a}_{1} }{\bar{a}_{3} } ; \, \, a_{2} =\frac{\bar{a}_{2} }{\bar{a}_{3} } ; \nonumber \\ \, \, & & \eta =\frac{X^{2} }{\varepsilon ^{2} } \frac{\bar{\eta }}{T} =\bar{\eta }\frac{m\rho }{\varepsilon ^{2} } .\, \,  \end{eqnarray}

When the order parameter equals either the minimum $a_{1} $, or the minimum $a_{3} $ of the potential polynomial the mobility \eqref{2.2} becomes singular only in the limit $\xi \to 0$. On the other hand, the mobility $M_{s} $ has equal minimal values $\frac{1}{\xi } $ at the stationary states $a_{1} ,a_{3} $ and the maximum at the intermediate one, $u_{m} =\frac{1}{2} \left(a_{1} +a_{3} \right)$. This maximum becomes singular only in the limit $\xi \to \xi _{s} =\theta \frac{1}{4} \left(a_{3} -a_{1} \right)^{2} $. It is also convenient to introduce notations
\begin{equation} \label{2.5)} a_{1} +a_{2} +a_{3} =\sigma ;\, \, \, \, a_{1} a_{2} +a_{2} a_{3} +a_{3} a_{1} =\zeta \, .   \end{equation}

Looking for the traveling-wave solution we introduce $z=x-vt$; then \eqref{2.1} takes the form
\begin{eqnarray} \label{2.6} & & -\alpha \frac{d}{dz} \left(\frac{v}{\alpha } u+u^{2} \right) \\ \nonumber & & =\frac{d}{dz} \left\{M_{i,s} \frac{d}{dz} \left[u^{3} -\sigma u^{2} +\zeta u-\frac{d^{2} u}{dz^{2} } -v\eta \frac{du}{dz} \right]\right\} .  \end{eqnarray}
Here we presumed $\alpha \ne 0$. Integrating \eqref{2.6} once, we get
\begin{eqnarray} \label{2.7} & & -\alpha \left(u^{2} +\frac{v}{\alpha } u+C\right) \\ \nonumber & &=M_{i,s} \frac{d}{dz} \left[u^{3} -\sigma u^{2} +\zeta u-\frac{d^{2} u}{dz^{2} } -v\eta \frac{du}{dz} \right] .  \end{eqnarray}
Here $C$ is an integration constant. We are looking for solution, which approaches values $u_{1} ,\, u_{2} $ at $\pm \infty $, $u_{1} <u_{2} $. For such a solution the simplest proper Ansatz is
\begin{equation} \label{2.8} \frac{du}{dz} =\kappa \left(u-u_{1} \right)\left(u-u_{2} \right)=\kappa \left(u^{2} -pu+q\right) .  \end{equation}
Here we denoted $p=u_{1} +u_{2} ,\, \, q=u_{1} u_{2} $; $\kappa $ is presently an unknown constant. The right-hand side of \eqref{2.7} is zero at $\pm \infty $. This means that $u_{1} ,\, u_{2} $ should be the roots of the quadratic polynomial in the left-hand side of \eqref{2.7}, i.e.,
\begin{equation} \label{2.9} p=-\frac{v}{\alpha } , \, \, \, C=q ,  \end{equation}
and Eq. \eqref{2.7} becomes
\begin{equation} \label{2.10} -\frac{\alpha }{\kappa } \frac{du}{dz} =M_{i,s} \frac{d}{dz} \left[u^{3} -\sigma u^{2} +\zeta u-\frac{d^{2} u}{dz^{2} } -v\eta \frac{du}{dz} \right] .  \end{equation}

Substituting \eqref{2.2} for $M_{i} $, we rewrite the latter equation as
\begin{eqnarray} \label{2.11} & &-\frac{\alpha }{\kappa } \left[\theta \left(u^{2} -2a_{i} u+a_{i}^{2} \right)+\xi \right]\frac{du}{dz} \nonumber \\ & &=\frac{d}{dz} \left[u^{3} -\sigma u^{2} +\zeta u-\frac{d^{2} u}{dz^{2} } -v\eta \frac{du}{dz} \right] . \end{eqnarray}
Substituting \eqref{2.3} for $M_{s} $, we rewrite \eqref{2.10} as
\begin{eqnarray} \label{2.12} & &-\frac{\alpha }{\kappa } \left[\theta \left(u^{2} -su+r\right)+\xi \right]\frac{du}{dz} \nonumber \\ & & =\frac{d}{dz} \left[u^{3} -\sigma u^{2} +\zeta u-\frac{d^{2} u}{dz^{2} } -v\eta \frac{du}{dz} \right] . \end{eqnarray}
Here we introduced for brevity $s=a_{1} +a_{3} , \, \, r=a_{1} a_{3} $. The only difference between \eqref{2.11} and \eqref{2.12} is that $s,\, r$ in \eqref{2.12} are changed to $2a_{i} ,\, a_{i}^{2} $. So we do calculations for \eqref{2.12} and present separately the final expressions corresponding to \eqref{2.11}. Using \eqref{2.8}, the second derivative is easily calculated:
\begin{equation} \label{2.13)} \frac{d^{2} u}{dz^{2} } =\kappa ^{2} \left[2u^{3} -3pu^{2} +\left(2q+p^{2} \right)u-pq\right] .  \end{equation}

Then the expression under the derivative in the right-hand side of \eqref{2.11} becomes
\begin{eqnarray} \label{2.14)} & & {u^{3} -\sigma u^{2} +\zeta u-\frac{d^{2} u}{dz^{2} } -v\eta \frac{du}{dz}} \nonumber \\ & & {=\left(1-2\kappa ^{2} \right)u^{3} +}{\left(3\kappa ^{2} p-v\eta \kappa -\sigma \right)u^{2}} \\ & & {+\left[-\kappa ^{2} \left(2q+p^{2} \right)+v\eta \kappa p+\zeta \right]u+\left(\kappa ^{2} pq-v\eta \kappa q\right)}  . \nonumber  \end{eqnarray}
Substitution of the latter expression into \eqref{2.12} yields
\begin{eqnarray} \label{2.15} & & {\left[-\alpha \theta u^{2} +\alpha \theta su-\alpha \left(\theta r+\xi \right)\right]\frac{du}{dz}}\nonumber \\ & & { =\left\{3\kappa \left(1-2\kappa ^{2} \right)u^{2} +2\kappa \left(3\kappa ^{2} p-v\eta \kappa -\sigma \right)u \right.} \\ & & {\left.+\kappa \left[-\kappa ^{2} \left(2q+p^{2} \right)+v\eta \kappa p+\zeta \right]\right\}\frac{du}{dz} } . \nonumber   \end{eqnarray}
The zero derivative, $\frac{du}{dz} =0$, corresponds to a constant solution; presuming $\frac{du}{dz} \ne 0$, we rewrite \eqref{2.15} as
\begin{eqnarray} \label{2.16)} & & {\left[3\kappa \left(1\!-\!2\kappa ^{2} \right)+\alpha \theta \right]u^{2} +\left[2\kappa \left(3\kappa ^{2} p\!-\!v\eta \kappa \!-\!\sigma \right)\!-\!\alpha \theta s\right]u}\nonumber \\ & & {+\!\left\{\kappa \left[\!-\!\kappa ^{2} \left(2q+p^{2} \right)\!+\!v\eta \kappa p+\zeta \right]\!+\!\alpha \left(\theta r+\xi \right)\right\}=0}  .  \end{eqnarray}

The latter equality should be satisfied identically; this yields three constraints. The first constraint is the same both for $M_{i} $ and $M_{s} $ cases :
\begin{equation} \label{2.17} \kappa ^{3} -\frac{1}{2} \kappa -\frac{1}{6} \alpha \theta =0 .  \end{equation}
The two other constraints for the $M_{s} $ case are
\begin{equation} \label{2.18} 3\kappa ^{3} p-v\eta \kappa ^{2} -\sigma \kappa -\frac{1}{2} \alpha \theta s=0,  \end{equation}
\begin{equation} \label{2.19} -\kappa ^{3} \left(2q+p^{2} \right)+v\eta \kappa ^{2} p+\left[\zeta \kappa +\alpha \left(\theta r+\xi \right)\right]=0 .  \end{equation}

Substitution of $2a_{i} ,\, a_{i}^{2} $ for $s,\, r$ yields the constraints for the case $M_{i} \, \, i=1,3$. We obtained three constraints \eqref{2.17}-\eqref{2.19}, imposed on the parameters of the system and parameters of the solution. If these constraints are satisfied, the solutions of Eq.\eqref{2.8} are simultaneously solutions of Eq.\eqref{2.6}. Integrating and taking the position of the maximal value of the modulus of the derivative $\frac{du}{dz} $ to be at $z=0$, we get
\begin{equation} \label{2.20} u=\frac{u_{2} +u_{1} }{2} -\frac{u_{2} -u_{1} }{2} \tanh\left(\frac{1}{2} \kappa \left(u_{2} -u_{1} \right)\left(x-vt\right)\right) .  \end{equation}
The dependence of the parameters of solution on the parameters of the system is analyzed in the next Section.

\setcounter{equation}{0}
\section{Parametric dependence of solution}\label{s3}

As it was shown in the previous Section~\ref{s2}, for Eq.\eqref{2.20} to be the solution of Eq.~\eqref{2.1} three constraints should be satisfied; there are three unknowns $\kappa \, $ and $u_{1} ,\, u_{2} $ (or $p,\, q$). Eq.\eqref{2.17} is a depressed cubic equation \cite{46}. This equation has one real root and two complex conjugate roots if $Q>0$, or three real roots, if $Q<0$, where $Q$ for Eq.\eqref{2.17} is
\begin{equation} \label{3.1)} Q=\frac{1}{4} \left(\frac{1}{6} \right)^{2} \left[\left(\alpha \theta \right)^{2} -\frac{2}{3} \right] .  \end{equation}
So, there are three possible values of $\kappa $ for $\left(\alpha \theta \right)^{2} <\frac{2}{3} $, and only a single value for $\left(\alpha \theta \right)^{2} >\frac{2}{3} $. For the latter case the real root of Eq.\eqref{2.17} can be written in the Cardano form [46], see Appendix~1, Eq.\eqref{A1.1}. However, if we are interested in the essential influence of the non-constancy of mobility, assuming $\left|\alpha \right|\theta \gg 1$ we get from Eq.\eqref{A1.1}
\begin{equation} \label{3.2)} \kappa \simeq \left(\frac{1}{6} \alpha \theta \right)^{\frac{1}{3} } +\left(\frac{1}{36\alpha \theta } \right)^{\frac{1}{3} } \, .   \end{equation}
On the other hand, for $\left|\alpha \right|\theta \ll 1$ the direct perturbation yields, see Appendix~1.
\begin{equation} \label{3.3} \kappa _{1,2} =\pm \frac{1}{\sqrt{2} } +\frac{1}{6} \alpha \theta ;\, \, \, \, \, \kappa _{3} =-\frac{1}{3} \alpha \theta  .  \end{equation}

In the limit $\theta \to 0$ the coefficients $\kappa _{1,2} $ given by \eqref{3.3} coincide with the values for the constant mobility \cite{2}, and $\kappa _{3} $ corresponds to the trivial constant solution. Considering $\kappa $ as given, Eqs.\eqref{2.18},\eqref{2.19} are the system to find $q$ and $p$ (or $u_{1} ,\, u_{2} $). Using Eq.\eqref{2.9} to eliminate $v$ from \eqref{2.18}, we get
\begin{equation} \label{3.4} \left(3\kappa ^{3} +\kappa ^{2} \alpha \eta \right)p=\kappa \sigma +\frac{1}{2} \alpha \theta s .  \end{equation}

If $\theta =0$ the latter equation coincides with corresponding constraint in \cite{2} (the $\kappa $ in \cite{2} was defined with the opposite sign). If additionally the symmetry constraint $\sigma =0$ on the stationary states of the potential was imposed, the condition $3\kappa +\alpha \eta =0$, i.e., the balance of ``forcing'' and dissipation, was the necessary condition for the existence of the traveling wave solution. In the present work $\theta \ne 0$ is essential, so we will always presume $3\kappa +\alpha \eta \ne 0$, and rewrite Eq. \eqref{3.4} as
\begin{equation} \label{3.5)} p=\frac{\kappa \sigma +\frac{1}{2} \alpha \theta s}{3\kappa ^{3} +\kappa ^{2} \alpha \eta }  .  \end{equation}
Then the velocity of the front is
\begin{equation} \label{3.6)} v=-\, \frac{\kappa \alpha \sigma +\frac{1}{2} \alpha ^{2} \theta s}{3\kappa ^{3} +\kappa ^{2} \alpha \eta }  .  \end{equation}

Remarkably, even if the symmetry constraint $\sigma =0$ is imposed, the velocity is non-zero. Considering the monotone mobility \eqref{2.2}, we set $s=2a_{i} $. If $i=1$, i.e., the mobility \eqref{2.2} is the decreasing function of $u$ in the interval $\left(a_{1} ,\, a_{3} \right)$, the absolute value of the velocity is smaller, than for the $i=3$ case, when the mobility is increasing. If $s=a_{1} +a_{3} $, i.e., for the non-monotone mobility \eqref{2.3}, the velocity is the arithmetic mean of the velocities for the above cases. Elimination of $v$ from \eqref{2.19} yields
\begin{equation} \label{3.7)} q=-\frac{1}{2} p^{2} \left(1+\frac{1}{\kappa } \alpha \eta \right)+\frac{1}{2\kappa ^{3} } \left[\alpha \left(\theta r+\xi \right)+\zeta \kappa \right] .  \end{equation}
The values of the order parameter at $\pm \infty $ are the roots of quadratic equation $u^{2} -pu+q=0$; the discriminant $R$ of this equation is
\begin{eqnarray} \label{3.8)} & &R=\frac{1}{4} p^{2} -q \\ & & =\left(\frac{\kappa \sigma +\frac{1}{2} \alpha \theta s}{3\kappa ^{3} +\kappa ^{2} \alpha \eta } \right)^{2} \left(\frac{3}{4} +\frac{\alpha \eta }{2\kappa } \right)-\frac{1}{2\kappa ^{3} } \left[\alpha \left(\theta r+\xi \right)+\zeta \kappa \right] . \nonumber  \end{eqnarray}

So the only necessary condition for the solution to exist is the existence of the real $u_{1} ,\, u_{2} $, i.e., $R>0$. For the weak dependence of mobility on the order parameter, i.e., $\left|\alpha \right|\theta \ll 1$, the positivity of $R$ in the zero order is checked in \cite{2}. So it is interesting to check the strong dependence of mobility on the order parameter, i.e., $\left|\alpha \right|\theta \gg 1$. In zero order one has $\kappa \simeq \left(\frac{1}{6} \alpha \theta \right)^{\frac{1}{3} } $. If we consider first the symmetric potential $\sigma =0$ and relatively weak viscosity $\eta $ and convective term $\alpha $ when $\frac{\alpha \eta }{\kappa } =\left(6\frac{\alpha ^{2} }{\theta } \right)^{\frac{1}{3} } \eta \ll 1$, the condition $R>0$ reduces to  $s^{2} >4r$. While for mobility $M_{s} $ this inequality, $\, \left(a_{1} +a_{3} \right)^{2} >4a_{1} a_{3} $, is always satisfied, for mobility $M_{i} $ it means $R=0$, that is, zero amplitude of the wave. In other words, for the latter (monotonic) case in the limit of large $\theta $ the non-zero solution is possible for the asymmetric potential, $\sigma \ne 0$, only.

On the other hand, if $1\ll \left|\alpha \right|\theta \ll \left|\alpha \right|^{3} \eta ^{3} $, for $\sigma =0$ $R$ is definitely negative. So for the symmetric potential if the viscosity and forcing (convection term) are strong enough, the amplitude of the solution approaches zero even for large $\theta $.

\setcounter{equation}{0}
\section{Viscous Cahn-Hilliard equation with the monotone singular mobility}\label{s4}

In Section~\ref{s2} we considered the convective-viscous CH equation \eqref{1.1} with $\alpha \ne 0$ and mobilities \eqref{1.2},\eqref{1.3}. In the present Section we consider special case $\alpha =0$, i.e., the viscous CH equation, however, with the singular mobilities \eqref{1.9},\eqref{1.10}:
\begin{eqnarray} \label{4.1}  \frac{\partial w}{\partial \bar{t}}& & =\frac{\partial }{\partial \bar{x}} \left\{\bar{K}_{i} \left(w\right)\frac{\partial }{\partial \bar{x}} \left[\rho \left(w-\bar{a}_{1} \right)\left(w-\bar{a}_{2} \right)\left(w-\bar{a}_{3} \right)\right.\right. \nonumber \\ & & \left.\left. -\varepsilon ^{2} \frac{\partial ^{2} w}{\partial \bar{x}^{2} } +\bar{\eta }\frac{\partial w}{\partial \bar{t}} \right]\right\}, \end{eqnarray}
\begin{equation} \label{4.2} \bar{K}_{i} \left(w\right)=\frac{\delta _{i} m}{\left(w-\bar{a}_{i} \right)} ;\, \, \, \, i=1,3;\, \, \, \delta _{1} =1;\, \, \delta _{3} =-1 .\end{equation}
We rewrite \eqref{4.1},\eqref{4.2} in nondimensional form; nearly all characteristic values and nondimensional notations remain the same; the only change is the characteristic time, which is now $\, T=\frac{\varepsilon ^{2} }{m\rho ^{2} \bar{a}_{3}^{3} } $:
\begin{eqnarray} \label{4.3} \frac{\partial u}{\partial t}& & =\frac{\partial }{\partial x} \left\{\bar{K}_{i} \frac{\partial }{\partial x} \left[\left(u-a_{1} \right)\left(u-a_{2} \right)\left(u-a_{3} \right) \right.\right. \nonumber \\ & & \left.\left.-\frac{\partial ^{2} u}{\partial x^{2} } +\eta \frac{\partial u}{\partial t} \right]\right\} ,  \end{eqnarray}
\begin{equation} \label{4.4)} K_{i} \left(u\right)=\frac{\delta _{i} }{\left(u-a_{i} \right)} ;\, \, \, \, i=1,3;\, \, \, \delta _{1} =1;\, \, \delta _{3} =-1 .  \end{equation}

Looking for the traveling-wave solution we transform \eqref{4.3} into
\begin{equation} \label{4.5)} -v\frac{du}{dz} =\frac{d}{dz} \left[K_{i} \frac{d}{dz} \left(u^{3} -\sigma u^{2} +\zeta u-\frac{d^{2} u}{dz^{2} } -v\eta \frac{du}{dz} \right)\right] .  \end{equation}
Integrating once, we get
\begin{eqnarray} \label{4.6} & &-v\delta _{i} \left(u-C\right)\left(u-a_{i} \right)\nonumber \\ & &=\frac{d}{dz} \left(u^{3} -\sigma u^{2} +\zeta u-\frac{d^{2} u}{dz^{2} } -v\eta \frac{du}{dz} \right) .  \end{eqnarray}
The solution of this equation must necessarily approach $a_{i} $ at $+\infty $ (or $-\infty $) and some presently unknown $u_{j} ,\, \, j\ne i$ at  $-\infty $ (or $+\infty $). We presume for definiteness $u_{1} <u_{3} $. Then the proper Ansatz for the solution is
\begin{eqnarray} \label{4.7} & &\frac{du}{dz} =\kappa \left(u-a_{i} \right)\left(u-u_{j} \right)\nonumber \\ & &=\kappa \left(u^{2} -pu+q\right);\, \, \, i\ne j;\, \, i,j=1,3 .  \end{eqnarray}
Here we denote $p=a_{i} +u_{j} ;\, \, q=a_{i} u_{j} $. Setting $C=u_{j} $ we rewrite \eqref{4.6} as
\begin{equation} \label{4.8} -\frac{v\delta _{i} }{\kappa } \frac{du}{dz} =\frac{d}{dz} \left(u^{3} -\sigma u^{2} +\zeta u-\frac{d^{2} u}{dz^{2} } -v\eta \frac{du}{dz} \right) .  \end{equation}

The expression under derivative in the right-hand side of the latter equation is again
\begin{eqnarray} \label{4.9)} & & {u^{3} -\sigma u^{2} +\zeta u-\frac{d^{2} u}{dz^{2} } -v\eta \frac{du}{dz}} \nonumber \\ & & =\left(1-2\kappa ^{2} \right)u^{3} +} {\left(3\kappa ^{2} p-v\eta \kappa -\sigma \right)u^{2} \\ & & {+\left[-\kappa ^{2} \left(2q+p^{2} \right)+v\eta \kappa p+\zeta \right]u+\left(\kappa ^{2} pq-v\eta \kappa q\right)} .  \nonumber  \end{eqnarray}
Substitution of the latter expression into \eqref{4.8} yields after integration
\begin{eqnarray} \label{4.10} & & {\left(1-2\kappa ^{2} \right)u^{3} +\left(3\kappa ^{2} p-v\eta \kappa -\sigma \right)u^{2}} \nonumber \\ & & { +\left[-\kappa ^{2} \left(2q+p^{2} \right)+v\eta \kappa p+\frac{v\delta _{i} }{\kappa } +\zeta \right]u} \\ & &{+\left(\kappa ^{2} pq-v\eta \kappa q\right)+C_{1} =0} . \nonumber   \end{eqnarray}

First we use the integration constant $C_{1} $ to cancel $u$-independent terms. Then the condition of identical fulfillment of Eq. \eqref{4.10} yields three constraints, imposed on three unknowns $\kappa ,\, u_{j} ,\, v$:
\begin{equation} \label{4.11)} 2\kappa ^{2} =1 ,  \end{equation}
\begin{equation} \label{4.12} 3\kappa ^{2} p-v\eta \kappa -\sigma =0 ,  \end{equation}
\begin{equation} \label{4.13} -\kappa ^{2} \left(2q+p^{2} \right)+v\eta \kappa p+\frac{v\delta _{i} }{\kappa } +\zeta =0 .  \end{equation}

Let us first consider the zero viscosity case. Then from Eq. \eqref{4.12} the  parameter$p$, i.e., $u_{j} $ is found; naturally, this determines $q$ also,
\begin{equation} \label{4.14} p=a_{i} +u_{j} =\frac{2}{3} \sigma  ,  \end{equation}
\begin{equation} \label{4.15} q=a_{i} u_{j} =a_{i} \left(\frac{2}{3} \sigma -a_{i} \right) .  \end{equation}
From Eqs. \eqref{4.13}-\eqref{4.15} the velocity is found
\begin{equation} \label{4.16)} v\delta _{i} =\kappa \left[a_{i} \left(\frac{2}{3} \sigma -a_{i} \right)+\frac{1}{2} \left(\frac{2}{3} \sigma \right)^{2} -\zeta \right] .\end{equation}

If $\eta \ne 0$
\begin{equation} \label{4.17)} v=\frac{1}{\eta \kappa } \left(\frac{3}{2} p-\sigma \right) \, \, \, .  \end{equation}
Substitution of the latter expression into \eqref{4.13} yields the quadratic equation for $p$
\begin{equation} \label{4.18)} p^{2} +\left(3\frac{\delta _{i} }{\eta } -\sigma -a_{i} \right)p-2\frac{\delta _{i} }{\eta } \sigma +\zeta +a_{i}^{2} =0 . \end{equation}
From $\sigma =0$ we have (see Eq. (2.5)) $\zeta =-\left(a_{1}^{2} +a_{3}^{2} +a_{1} a_{3} \right)$ and the latter equation has real roots, i.e., traveling wave solutions exist even for the symmetric potential.

\setcounter{equation}{0}
\section{Discussion}\label{s5}

In the present work we considered several exactly solvable modifications of the convective-viscous and viscous CH equation with the order-parameter-dependence of mobility. Generally, the dependence of mobility in the CH equation on the order parameter was evident already from the original derivation \cite{4}. However, explicit introduction of such dependence results in tremendous calculation difficulties, so the constant mobility approximation was commonly used. On the other hand, in several cases this dependence appeared to be crucial, so more realistic expressions were applied. Most commonly, the positive powers or polynomial of the order parameter were used. For such dependences only approximate and numeric solutions were found.

On the other hand, in several fields the mobilities (or diffusivities) with negative powers of the order parameter (concentration) were found to be a proper approximation. We consider several mobilities with this type of functional dependences on the order parameter. We have found the exact traveling wave solutions with three types of mobility for the convective-viscous CH equation, and with two different types of mobility for the viscous CH equation. Remarkably, some of these expressions for the mobility appear to be rather good approximation to the positive-power expressions, of course in a limited interval of the parameters, see Appendix~2. This approximation fails in the case of the degenerated in the equilibrium states mobilities. However it is worth to mention that the degeneracy of mobility in the equilibrium state is based on the following simple qualitative explanation. If the diffusion in a binary system proceeds via the exchange of positions, say, between A and B atoms, it stops in pure A and pure B phases. However if both the initial and final phase contain both A and B, but with different stoichiometry, this will not be the case; then a reciprocal quadratic dependence of the mobility on the order parameter can be a proper approximation.

The non-constancy of mobility influences essentially the velocity of the transition front: there is non-zero velocity even for the symmetric potential and without any special balance between forcing (convection) and dissipation, as in the constant mobility case. There is a qualitative difference of the solutions for the reciprocal linear and reciprocal quadratic mobilities. While for the reciprocal linear mobility the steepness of the front can be positive or negative, like in the constant mobility case, for the reciprocal quadratic mobility its sign coincides with the sign of the convective term.

The given examples do not exhaust the forms of mobilities which permit exact solutions for the convective-viscous CH equation; some of them will be given in the following communications.

\setcounter{equation}{0}
\subsection*{Appendix 1} \label{s:A1}
\renewcommand{\theequation}{A1.\arabic{equation}}

For $\left(\alpha \theta \right)^{2} >\frac{2}{3} $ the single real root of Eq.\eqref{2.17} can be written in the Cardano form \cite{46}:
\begin{eqnarray} \label{A1.1} & &\kappa =\left(\frac{1}{12} \right)^{\frac{1}{3} } \\ \nonumber & &\times \left\{\left[\alpha \theta +\sqrt{\left(\alpha \theta \right)^{2} -\frac{2}{3} } \right]^{\frac{1}{3} } \right.   +\left. \left[\alpha \theta -\sqrt{\left(\alpha \theta \right)^{2} -\frac{2}{3} } \right]^{\frac{1}{3} } \right\}. \end{eqnarray}
However, if we are interested in the essential influence of the non-constancy of mobility, i.e., $\left|\alpha \right|\theta \gg 1$, the latter expression is simplified to
\begin{equation} \label{A1.2)} \kappa \simeq \left(\frac{1}{6} \alpha \theta \right)^{\frac{1}{3} } +\left(\frac{1}{36\alpha \theta } \right)^{\frac{1}{3} } ,  \, \, \, \, \left|\alpha \right|\theta \gg 1 . \end{equation}
On the other hand, if $\left|\alpha \right|\theta \ll 1$ the direct perturbation
\begin{equation} \label{A1.3)} \kappa =\kappa _{0} +\left(\alpha \theta \right)\kappa _{1} +O\left(\left(\alpha \theta \right)^{2} \right) \end{equation}
yields
\begin{equation} \label{A1.4} \kappa _{0}^{3} -\frac{1}{2} \kappa _{0} =0 \, ,   \end{equation}
\begin{equation} \label{A1.5)} \, \left(\, 3\kappa _{0}^{2} -\frac{1}{2} \right)\kappa _{1} =\frac{1}{6} . \end{equation}
Eq.\eqref{A1.4} yields three possible values: $\kappa _{01} =\frac{1}{\sqrt{2} } ;\, \, \kappa _{02} =-\frac{1}{\sqrt{2} } ;\, \, \kappa _{03} =0$. Correspondingly, within the first order corrections there are three values of $\kappa $
\begin{equation} \label{A1.6)} \kappa _{1,2} =\pm \frac{1}{\sqrt{2} } +\frac{1}{6} \alpha \theta , \, \, \, \, \, \kappa _{3} =-\frac{1}{3} \alpha \theta  .  \end{equation}


\setcounter{equation}{0}
\subsection*{Appendix 2} \label{s:A2}
\renewcommand{\theequation}{A2.\arabic{equation}}

Here we compare the most common expression for the mobility
\begin{equation} \label{A2.1} M_{c} =-\theta \left(u-a_{1} \right)\left(u-a_{3} \right)+\frac{1}{\xi }  \end{equation}
with our expression \eqref{2.3},
\begin{equation} \label{A2.2} M_{s} =\frac{1}{\theta \left(u-a_{1} \right)\left(u-a_{3} \right)+\xi } . \end{equation}
We have selected the mobilities \eqref{A2.1},\eqref{A2.2} to have the same value $\frac{1}{\xi } $ at stationary states of the potential. For $a_{1} \le u\le a_{3} $ the maximal values of these mobilities are
\begin{equation} \label{A2.3)} \max M_{c} =\frac{1}{4} \theta \left(a_{3} -a_{1} \right)^{2} +\frac{1}{\xi } , \end{equation}
\begin{equation} \label{A2.4)} \max M_{s} =\frac{1}{-\frac{1}{4} \theta \left(a_{3} -a_{1} \right)^{2} +\xi } . \end{equation}
For the maximal values to coincide it should be
\begin{equation} \label{A2.5)} \theta \left(a_{3} -a_{1} \right)^{2} =4\frac{\xi ^{2} -1}{\xi } . \end{equation}

\begin{figure}[t]
    \centering
    \begin{subfigure}[b]{0.35\textwidth}
        \includegraphics[width=1.2\linewidth]{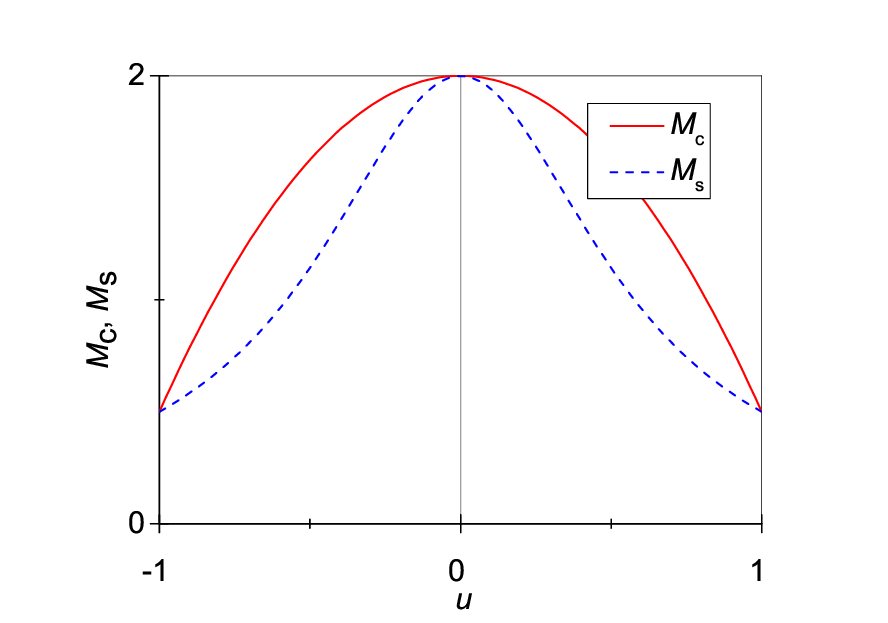}
        \caption{Parameter $\tau=0.5$}
    \end{subfigure}
    \begin{subfigure}[b]{0.35\textwidth}
        \includegraphics[width=1.2\linewidth]{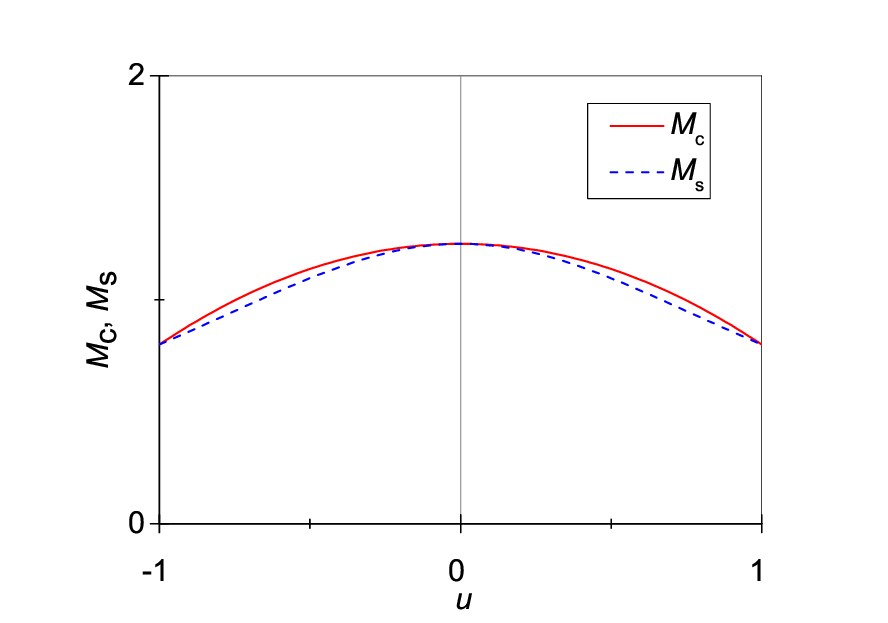}
        \caption{Parameter $\tau=0.8$}
    \end{subfigure}
    \caption{Dependence of mobilities $M_c$ and $M_s$ on parameter $\tau$ in the interval $-1\le u\le 1$ of the order paramener $u$}\label{Fig:1}
\end{figure}

Taking for simplicity $a_{1} =-1$ and remembering that per definition $a_{3} =1$, we get $\theta =\frac{\xi ^{2} -1}{\xi } $. We remind that it is necessary that $\xi >\xi _{s} $ and that for the present values of $a_{i} $ one has $\xi _{s} =\theta $,  i.e., $\xi >\theta $. So we can write $\xi =\theta +\tau $, where $\tau $ is some positive number. Substitution of this expression for $\xi $ yields allowed values of $\theta $, and consequently of $\xi $,
\begin{equation} \label{A2.6)} \, \theta =\frac{1-\tau ^{2} }{\tau } , \, \, \, \, \xi =\frac{1}{\tau }  .  \end{equation}
Because $\theta >0$ only $\tau <1$ are allowed. Now we plot the mobilities \eqref{A2.1},\eqref{A2.2} on the interval $-1\le u\le 1$ for this parametric dependence (see Fig.~1)
\begin{equation} \label{A2.7)} M_{c} =\frac{1-\tau ^{2} }{\tau } \left(1-u^{2} \right)+\tau , \end{equation}
\begin{equation} \label{A2.8)} M_{s} =\frac{1}{\frac{1-\tau ^{2} }{\tau } \left(u^{2} -1\right)+\frac{1}{\tau } } =\frac{\tau }{1-\left(1-\tau ^{2} \right)\left(1-u^{2} \right)} . \end{equation}
Evidently, for $\tau $ close enough to unity the mobility $M_{s} $ is rather good approximation of $M_{c} $.

{\small \topsep 0.6ex

}

\end{document}